\documentclass[english,showpacs,floatfix,12pt]{revtex4}
\usepackage[dvips]{graphicx}
\usepackage{longtable}
\usepackage{amsmath,amssymb}
\usepackage{dcolumn}
\usepackage{latexsym}
\usepackage{babel}
\usepackage[latin1]{inputenc}
\usepackage{color}
\usepackage[colorlinks]{hyperref}
\def\al{\alpha}

\def\nb{\nabla}
\def\pa{\partial}

\def\ga{\gamma}
\def\be{\beta}
\def\ck{\check}
\def\dl{\delta}
\def\Dl{\Delta}

\def\lan{\langle}
\def\ran{\rangle}

\def\sg{\sigma}
\def\wt{\widetilde}
\def\nn{\nonumber}
\def\diag{\mbox {diag}}

\def\l{\left}
\def\r{\right}

\begin{document}
\title{\Large \bf Local Lorentz Transformation and Mass-Energy Relation of Spinor}
\author{Ying-Qiu Gu}\email{yqgu@fudan.edu.cn} \affiliation{School of
Mathematical Science, Fudan University, Shanghai 200433, China}
\pacs{ 03.65.Ta, 03.50.Kk, 03.65.Pm, 11.10.Lm}
\date{30th May 2017}
\begin{abstract}
In this paper, we strictly establish classical concepts and
relations according to a Dirac equation with scalar, vector and
nonlinear potentials. To calculate classical parameters for moving
spinor, the local Lorentz transformations for parameters are
derived. The calculation shows that different kinds of potentials
result in different energy-speed relations, and the energy-speed
relations for these potentials are derived in detail. The usual
mass-energy relation $E = mc^2$ holds only for the linear spinor.
The energy-speed relations can be used as fingerprints to identify
the interactive potentials of a particle by elaborated experiments.
The analysis and results of this paper can also provide some natural
explanations for the foundation of quantum mechanics, and clarify
some long-standing puzzles in the theory.

\vskip0.3cm\noindent {\bf Keywords: } {\em mass-energy relation,
classical approximation, local Lorentz transformation, foundation of
quantum mechanics}
\end{abstract} \maketitle

\section{Introduction}
\setcounter{equation}{0}

Einstein's mass-energy relation
\begin{eqnarray}E=mc^2,\qquad m={m_0}\l({1-\frac
{v^2}{c^2}}\r)^{-\frac 1 2}, \label{1.0}\end{eqnarray} is the
apotheosis of the elegance and simplicity of modern science, which
becomes a cornerstone of the modern physics. However its original
derivation is based on an idealized model and valid only for linear
theories\cite{E1}. For a spinor with self-potential, the
energy-speed relation includes fine structure\cite{gu1}, namely, the
energy caused by potentials differs from (\ref{1.0}). In $E=mc^2$,
the total mass $m$ should weakly depend on the speed $v$ of the
particle, and a different kind of potential leads to a different
energy-speed relation. So it is meaningful to derive the detailed
energy-speed relation for each kind potential. These relations can
be used as the fingerprints of the interactive potentials, and can
be measured by elaborated experiments.

This problem is also closely related with some important puzzles in
fundamental physics as listed in \cite{sml}, that is, the
consistence of general relativity and quantum theory, the foundation
of quantum mechanics, the unification for the particles and
interactions, dark matter and dark energy, and so on. Among these
puzzles the foundation of quantum mechanics is the key. From its
beginning, quantum mechanics provides us counterintuitive concepts
and pictures for the world such as the principle of the uncertainty,
wave function of probability, correspondence principle of operators
with parameters. These concepts have greatly changed our world
outlook. However, with its continuous and tremendous successes,
quantum mechanics also makes troubles continually for us.

To explain the foundation of quantum mechanics, we should clarify
the relationship between quantum mechanics with classical mechanics
at first. In the textbooks we have a standard procedure. By
correspondence principle, a classical parameter such as energy or
momentum corresponds to an operator, and  the operator acts on state
functions to form a dynamics in quantum mechanics. On the contrary,
by calculating the mean value of an operator we get the classical
approximation for the parameter. Such parallel relation easily
results in confusion in logic.

In this paper, we take a generalized Dirac equation with nonlinear,
scalar and vector interactive potentials as example to show the
local Lorentz transformation for classical parameters and derive
complete classical mechanics and energy-speed relations for the
spinor. These derivations and results imply a natural explanation
for the relationship between classical mechanics and quantum
mechanics. We find some puzzles may be avoidable. At last, a crucial
experiment is proposed.

\section{Local Lorentz transformation Laws for classical parameters}
\setcounter{equation}{0}

First, we introduce some notations and conventions. Taking the
Minkowski metric as $\eta_{\mu\nu}={\rm diag}(1,-1,-1,-1)$, and
$c=1$ as the unit of speed.  Pauli matrices are given by
\begin{eqnarray}
 {\vec\sg}=(\sg^{k})= \left \{\l(\begin{array}{cc}
 0 & 1 \\ 1 & 0 \end{array} \right),\l(\begin{array}{cc}
 0 & -i \\ i & 0\end{array}\r),\l(\begin{array}{cc}
 1 & 0 \\ 0 & -1 \end{array} \right)
 \right\}.\label{1.1}\end{eqnarray}
Instead of Dirac matrices $\ga^\mu$, define $4\times4$ Hermitian
matrices as follows for the convenience of calculation,
\begin{eqnarray}\al^\mu=\left\{\left ( \begin{array}{cc} I & 0 \\
0 & I \end{array} \right),\left (\begin{array}{cc} 0 & \vec\sg \\
\vec\sg & 0 \end{array}
\right)\right\},\qquad \ga =\left ( \begin{array}{cc} I & ~0 \\
0 & -I \end{array} \right).\label{1.2} \end{eqnarray}  We use Greek
characters stand for 4-vector index and Latin characters for spatial
index.

Considering the following nonlinear spinor equation
\begin{eqnarray}
\al^\mu (\hbar i\pa_\mu-eA_\mu)\phi = (\mu c-F')\ga\phi,
\label{*2.1}
\end{eqnarray}
where the coefficient $\mu>0$ is a constant mass, $F(\ck\ga)>0$ is
the nonlinear function of the quadratic scalar $\ck\ga\equiv
\phi^+\ga\phi$. Solutions to (\ref{*2.1}) have particle-wave
duality. Some numeric simulations were provided in
\cite{3}-\cite{gu3}.

For Dirac equation (\ref{*2.1}), we have current conservation law
due to the gauge invariance
\begin{eqnarray}\pa_\mu\rho^\mu=0, \qquad \rho^\mu= \phi^+\al^\mu\phi, \label{*2.5}\end{eqnarray}  which leads to the normalizing condition
\begin{eqnarray}\int_{R^3}\rho^0 d^3 x=1. \label{*2.7}\end{eqnarray}
Due to (\ref{*2.5}) and (\ref{*2.7}) , we can define classical
concepts for the spinor $\phi$.

{\bf Definition 1.} For spinor field  $\phi$, we define the {\bf
coordinate} $\vec X$ and {\bf velocity} $\vec v$ respectively by
\begin{eqnarray}\vec X(t)=\int_{R^3} \vec x \rho^0 d^3 x,\qquad \vec v=\frac d {dt} \vec
X,\label{*2.8} \end{eqnarray} where $t=x^0$. The coordinate system
with origin $\vec X=0$ is called {\bf the central coordinate system}
of the spinor.

{\bf Lemma 1.} {\em For a spinor, we have
\begin{eqnarray}
\vec v=\int_{R^3} \vec \rho d^3x.\label{*2.9}
\end{eqnarray}
}

{\bf Proof.} By the current conservation law (\ref{*2.5}), we have
\begin{eqnarray}
\vec v=\int_{R^3} \vec x\pa_0\rho^0 d^3 x=-\int_{R^3} \vec x
\nb\cdot \vec \rho d^3x =\int_{R^3} \vec \rho d^3x.\label{2.9*}
\end{eqnarray}

By (\ref{*2.7})  and (\ref{*2.9}), we have the classical
approximation of the spinor, i.e.

{\bf Definition 2.} if the moving scale of a spinor is much larger
than its mean radius $\bar r =\int |\vec x -\vec X|\rho^0 d^3 x$, we
have {\bf point-particle model} for the spinor,
\begin{eqnarray} \rho^\mu\to u^\mu
\sqrt{1-v^2}\dl(\vec x-\vec X), \label{*2.10} \end{eqnarray} where
$u^\mu$ is the 4-vector speed
\begin{eqnarray}u^\mu\equiv (\xi, \xi \vec v),
\qquad \xi=\frac 1 {\sqrt{1-v^2}}.\label{*2.11}
\end{eqnarray}

(\ref{*2.10}) is the origin of the classical models `mass-point' and
`point-charge'\cite{gu1,*gu7}. The classical variables such as
momentum and energy are also defined as the integrals of some
N\"other charges similar to (\ref{2.9*}). However, in usual cases,
the values depend on the spinor at energy eigen state or not, due to
absence of conservation law similar to (\ref{*2.5}) for the spinor
as a closed system. So it is necessary to distinguish the different
states of the spinor.

{\bf Definition 3.} If a spinor takes energy eigenstate in the
central coordinate system and moves smoothly without emitting and
absorbing energy quantum, we call it is at the {\bf particle state}.
Otherwise, the spinor is in the process of exchanging energy with
its environment, we call it in the {\bf quantum process}.

For a spinor at particle state moving smoothly, we can clearly
define the classical parameters such as ``momentum", ``energy" and
``mass", and then derive the classical mechanics from (\ref{*2.1}).
But for the spinor in the quantum process, the detailed description
of the process should be the original equation (\ref{*2.1}) and
quantum theory.

In the usual case, we can calculate only the classical parameters in
central coordinate system, i.e. the proper parameters. In a general
coordinate system, we should derive the moving parameters by local
Lorentz transformation.  Since the rotational transformation is
trivial, we consider only the boosting one. Considering the central
coordinate system of the spinor with coordinate $\bar x^\mu$, which
moves along $x^1$ at speed $v$, and $\bar x^k(k\ne0)$ is parallel to
$x^k$, so $\bar x^k=0$ corresponds to the mass center $\vec X^k(t)$
of the field $\phi$. Then the Lorentz transformation between $x^\mu$
and $\bar x^\mu$ in the form of matrix is given by
\begin{eqnarray}x=L(v)\bar x,\qquad \bar x=L(v)^{-1} x=L(-v)
x\label{*2.12} \end{eqnarray} where $x=(t, x^1, x^2, x^3)^T$, $\bar
x=(\bar x^0, \bar x^1, \bar x^2, \bar x^3)^T$ and
\begin{eqnarray}
L(v)=\diag \left(\l( \begin{array}{cc}
 \xi & \xi v \cr \xi v & \xi\end{array} \right),1,1\right)=(L^\mu_{~\nu}).\label{*2.13} \end{eqnarray}

Assuming $S, P^\mu$ and $T^{\mu\nu}$ are any scalar, vector and
tensor defined by some real functions of $\phi$ and their
derivatives such as $S=\ck\ga^2$, $T^{\mu\nu}=\Re\lan \phi^+\al^\mu
i\pa^\nu\phi\ran$ etc., where $\Re\lan\ran$ stands for taking real
part. For a spinor at particle state, all these functions in central
coordinate system are independent of proper time $\bar x^0$. So the
spatial integrals of these functions define the proper classical
parameters of the spinor, which are all constants. Their Lorentz
transformation laws are given by

{\bf Theorem 2.} {\em For a spinor at particle state, the integrals
of covariant functions $S, P^\mu$ and $T^{\mu\nu}$ satisfy the
following instantaneous Lorentz transformation laws under the
boosting transformation (\ref{*2.12}) between $x^\mu$ and $\bar
x^\mu$ at $t\equiv t_0$ or $d t=0$,
\begin{eqnarray}
I &\equiv& \int_{R^3} S(x) d^3x = \sqrt{1-v^2} \bar I, \label{*2.14}\\
I^\mu &\equiv& \int_{R^3} P^\mu(x) d^3x = \sqrt{1-v^2}
L^\mu_{~\nu} \bar I^\nu, \label{*2.15}\\
I^{\mu\nu}&\equiv& \int_{R^3} T^{\mu\nu}(x) d^3x =
\sqrt{1-v^2}L^\mu_{~\al}L^\nu_{~\be}\bar I^{\al\be}, \label{*2.16}
\end{eqnarray} where $\bar I, \bar I^\mu, \bar I^{\mu\nu}$ are the proper parameters defined in the central coordinate system}
\begin{eqnarray}
\bar I =\int_{R^3} S(\bar x) d^3\bar x, \quad \bar I^\mu =
\int_{R^3} \bar P^\mu d^3\bar x, \quad \bar I^{\mu\nu}= \int_{R^3}
\bar T^{\mu\nu} d^3\bar x. \label{*2.17} \end{eqnarray}

{\bf Proof.} We take (\ref{*2.15}) as an example to show the proof.
For a spinor at the particle state, by the transformation law of
contravariant vector, we have
\begin{eqnarray}P^\mu (x)=L^\mu_{~\nu}\bar P^\nu (\bar x)
=L^\mu_{~\nu}\bar P^\nu (\bar x^1,\bar x^2,\bar x^3)=P^\mu
(\xi(x^1-v t),x^2,x^3). \label{*2.18} \end{eqnarray} So the integral
can be calculated as follows
\begin{eqnarray}
I^\mu &=& \int_{R^3} P^\mu (x) d^3x\left|_{d t=0}\right.\nn\\
& = & \int_{R^3} P^\mu (\xi(x^1-v
t),x^2,x^3) \sqrt{1-v^2}d[\xi(x^1-vt)]dx^2 dx^3\label{*2.19}  \\
&=& \int_{R^3} L^\mu_{~\nu}\bar  P^\nu (\bar x) \sqrt{1-v^2} d^3
\bar x=\sqrt{1-v^2} L^\mu_{~\nu}\bar I^\nu.\nn
\end{eqnarray} The proof is finished.

{For the above calculations, some explanations are given in the
following:

{\bf Remarks 1.} The integration-domain of
Eqs.(\ref{*2.14})-(\ref{*2.17}) is the realistic space of the world,
which is a simultaneous hypersurface $t\equiv t_0$ in one special
coordinate system. In Ref.\cite{sbt} we analyzed the existence and
uniqueness of this simultaneous hypersurface in the real world. In
the case of Minkowski space-time, the integration-domain becomes
$R^3$ in any Descartes coordinate system. However, a relativistic
factor $\sqrt{1-v^2}$ will appear in the integrals under Lorentz
transformation as derived in (\ref{*2.19}), because the space is
tilted and its measure is changed in the new coordinate system.}

{\bf Remarks 2.} { The Lorentz transformation laws (\ref{*2.14}), (\ref{*2.15}) and (\ref{*2.16}) are valid for varying speed $v(t)$, because the integrals are related only  to the simultaneous condition $dt=0$, and the relations are related only to algebraic calculations.} 
{Since spinor has only a tiny micro structure in contrast with the
space-time, the relations Eqs.(\ref{*2.14})-(\ref{*2.16}) actually
hold in the tangent space-time $\ga_\al \dl X^\al$ of curved
manifold with the spinor central coordinate $\vec X=0$. So we call
them local Lorentz transformation. Besides, when we consider
many-body problem, each spinor has a different velocity and then
involves a different Lorentz transformation.}

{\bf Remarks 3.} { When the spinor is not at the particle state, the
covariant integrands will depend on the proper time $\bar x^0$, so
the calculation (\ref{*2.19}) becomes approximate in general cases,
and consequently the relations (\ref{*2.14})-(\ref{*2.16}) usually
hold only approximately unless the integrand satisfies the
conservation law similar to (\ref{*2.5}).}

\section{The classical mechanics for a spinor and mass-energy Relation}
\setcounter{equation}{0}

In this section, we take the following more general Lagrangian as an
example to derive classical mechanics for a spinor,
\begin{eqnarray}
{\cal L}&=&\phi^+\al^\mu (\hbar i\pa_\mu-e A_{\mu})\phi-\mu\ck
\ga+F(\ck\ga)-s\ck\ga G \nn
\\ &~&-\frac 1 2 \pa_\mu A_\nu\pa^\mu A^\nu-\frac 1 2
(\pa_\mu G\pa^\mu G-b^2 G^2), \label{2.1}
\end{eqnarray} where $A^\mu$ and $G$ are the self-potential of the spinor,  $F(\ck\ga)>0$ is a concave function satisfying
\begin{eqnarray} F'(\ck\ga)\ck\ga>F(\ck\ga),\qquad ({\mbox{for}}~
\ck\ga>0).\label{2.2}\end{eqnarray} The corresponding dynamical
equation is given by
\begin{eqnarray}
\al^\mu(\hbar i\pa_\mu-e A_\mu)\phi &=&(\mu+s G-F')\ga\phi, \label{2.3}\\
\pa_\al\pa^\al  A^\mu &=& e\rho^\mu,
\label{2.4}\\
(\pa_\al\pa^\al +b^2) G~&=& s\ck\ga. \label{2.5}
\end{eqnarray}
The Hamiltonian form of (\ref{2.3}) reads
\begin{eqnarray}\hbar i \pa_t\phi=\hat H\phi,~~ \hat H=eA_0+\vec\al\cdot(-\hbar i\nb-e \vec
A)+(\mu+s G-F')\ga. \label{219} \end{eqnarray} The complete
dynamical equation of $A^\mu$ in 3-d form is given by the following
Maxwell equation\cite{gu1,quat}
\begin{eqnarray}
  \left \{\begin{array}{lll}
 &  \vec{E}=-\nb A^0-\pa_0\vec{A},& \qquad  \vec{B}=\nb \times \vec
{A},\\
 &  \nb\cdot\vec{E}=e\rho^0,& \qquad  \nb\times\vec{E}=-\pa_0\vec{B},\\
 &  \nb\cdot\vec{B}=0,& \qquad  \nb\times \vec{B}=\pa_0\vec{E}+ e\vec{\rho},
  \end{array}\right.\label{2.0}\end{eqnarray}
where $\vec A=(A^1,A^2,A^3)$ is the spatial part of a contravariant
vector $A^\mu$. For the scalar $G$ we have not similar decomposition
with manifest physical meanings.

Assuming the spinor is at particle state, we can define the
classical momentum and energy for the spinor according to N\"other's
theorem\cite{gu1,grn},

{\bf Definition 4.} Define the 4-vector {\bf momentum} $p^\mu$ and
{\bf energy} $E$ of the particle described by (\ref{2.1})
respectively by
\begin{eqnarray}
p^\mu&\equiv&\int_{R^3}\phi^{+}_{k}(\hbar i\pa^\mu-eA^\mu)\phi d^3x,\label{2.7} \\
E~ &\equiv& \int_{R^3}\left(\sum_{\forall f} \frac{\pa {\cal
L}}{\pa(\pa_t f)}\pa_t f-{\cal L}\right)d^3 x =p^0+E_F+E_A+E_G,
 \label{2.8}
\end{eqnarray}
in which
\begin{eqnarray}
E_F&=&\int_{R^3}(F'\ck\ga-F)d^3 x,\label{2.9}\\
E_A &=&-\frac 1 2  \int_{R^3} (\pa_0  {A}_\mu\pa^0  {A}^\mu
 +\nb  {A}_\mu\cdot\nb  {A}^\mu -2 e \rho_0  {A}^0)d^3
 x,\label{2.10}\\
 E_G&=&-\frac 1 2 \int_{R^3}(\pa_0
  {G}\pa^0  {G}+\nb  {G}\cdot\nb
  {G} +b^2  {G}^2)d^3 x, \label{2.11}
\end{eqnarray}
where we take the spinor as a closed system, and  $ {A}^\mu$ and $
{G}$ are self-potentials satisfying the natural boundary condition.

In the central coordinate system $\bar x^\mu$ of the spinor, since
the spinor takes the energy eigenstate, we have $\pa_0 {A}^\mu=\pa_0
{G}=0$. By the Green's functions of (\ref{2.4}) and (\ref{2.5}), we
can calculate the corresponding static energy as follows
\begin{eqnarray}
W_F&=&\int_{R^3}(F'\ck\ga-F)d^3 \bar x>0,\label{2.12} \\
W_A &=&\frac 1 2   \int_{R^3} ( {A}_\mu \Dl {A}^\mu+2  e \rho_0
{A}^0)d^3\bar x \nn \\  &=&\frac 1 2 e \int_{R^3} (\rho_0
 {A}^0+\vec \rho\cdot
\vec { {A}}) d^3\bar x \nn \\
 &=&\frac {  e^2}{8\pi}\int_{R^6} \frac 1 r
{[|\phi(\bar x)|^2|\phi(\bar y)|^2+\vec \rho (\bar x)\cdot\vec
\rho(\bar y)]}d^3 \bar x d^3\bar y\nn \\ &\dot=&\frac {
e^2}{8\pi}\int_{R^6}\frac 1 r {|\phi(\bar x)|^2|
\phi(\bar y)|^2}d^3 \bar x d^3\bar y,\label{2.13} \\
W_G &=& \frac 1 2 \int_{R^3}  {G}(\Dl -b^2)  {G} d^3\bar x
 =-\frac 1 2 s \int_{R^3} \ck\ga  {G} d^3\bar x \nn\\
 &=&-\frac {  s^2}{8\pi}\int_{R^6}\frac {\exp({-br})} r
 {\ck\ga(\bar x)\ck\ga(\bar y)}d^3\bar x d^3\bar y, \label{2.14}
\end{eqnarray}
where $r=|\bar x-\bar y|$.  { By (\ref{2.13}) and (\ref{2.14}), we
find that $W_A$ provides positive self-energy, but $W_G$ provides
negative one, so the scalar field is quite different from the vector
one in some aspects.}

According to theorem 2, making local Lorentz transformation, we get
energy-speed relation for each part in (\ref{2.8}) for a moving
spinor $\phi$ as follows,
\begin{eqnarray}
E_F&=&W_F\sqrt{1-v^2},\label{2.15}\\
E_A &\dot =&W_A \left(\frac 2{\sqrt{1-v ^2}}-\sqrt{1-v
^2}-\frac{2v^2 }
 {3\sqrt{1-v^2 }}\right),\label{2.16} \\
 E_G &=& W_G \left(\sqrt{1-v
^2}+\frac{2v^2 }
 {3\sqrt{1-v^2 }}\right)-W_b\frac {v^2}{\sqrt{1-v^2}}. \label{2.17}
\end{eqnarray}
where
\begin{eqnarray}
W_b=\frac  1 3 \left( \frac {bs}{4\pi} \right)^2
\int_{R^3}\left(\int_{R^3} \frac {\exp(-br)} r \ck\ga(\bar y)d^3
\bar y\right)^2 d^3 \bar x.\end{eqnarray}

Now we examine the term $p^\mu$. It is easy to check the following
Ehrenfest theorem.

{\bf Lemma 3.} {\em For any Hermitian operator $\hat P$ and
corresponding classical quantity $P$ for the spinor defined by
\begin{eqnarray}
P\equiv \int_{R^3}\phi^+\hat P\phi d^3x,\label{2.18}
\end{eqnarray} we have
\begin{eqnarray}
 \frac d {dt} P=\int_{R^3}
\phi^+\left(\pa_t \hat P+i[\hat H,\hat P]\right)\phi d^3
x,\label{2.19} \end{eqnarray} where $[\hat H,\hat P]=\hat H\hat
P-\hat P\hat H$}.

By Lemma 3 and the dynamical equation (\ref{219}), we have\cite{gu1}

{\bf Theorem 4.} {\em For 4-vector momentum $p^\mu$ defined by
(\ref{2.7}), we have the following rigorous dynamical equations
\begin{eqnarray}
\left\{
\begin{array}{lll} \frac d {dt}p^0 &=&\int(e\vec \rho\cdot \vec
E+s\ck\ga\pa_0G)d^3 x-\frac d{dt}E_F  ,\\
\frac d {dt}\vec p &=&\int [e(\rho^0\vec E+\vec \rho\times\vec
B)-s\ck\ga\nb G] d^3 x,
\end{array}\right. \label{2.20} \end{eqnarray}
where $\vec E$ and $\vec B$ include the intensity of external
potential of ${A}^\mu$.}

Substituting the classical approximation (\ref{*2.10}) into
(\ref{2.20}), we get the Newtonian second law for the spinor $\phi$
\begin{eqnarray}
\left\{
\begin{array}{lll} \frac d {dt}p^0 &=&e\vec v\cdot \vec
E(t,\vec X)+s W_\ga \sqrt{1-v^2} \pa_t G(t,\vec X)-\frac d{dt}E_F  ,\\
\frac d {dt}\vec p &=& e (\vec E+\vec v\times\vec B)-sW_\ga
\sqrt{1-v^2} \nb G(t,\vec X),
\end{array}\right. \label{220} \end{eqnarray}
where $W_\ga=\int_{R^3} \ck\ga d^3\bar x$. By the dynamical equation
(\ref{2.3}), we get\cite{gu1}
\begin{eqnarray}
 p^\mu=\int_{R^3}{\Re}\lan \phi^{+} \al^\mu(\hbar i
\pa_0-eA_0)\phi\ran d^3x.\label{2.21}\end{eqnarray} Assuming the
spinor takes the energy eigenstate and moves smoothly, by
(\ref{*2.10}) and (\ref{2.21}) we find $p^\mu \propto u^\mu$, then
by covariance we have
\begin{eqnarray}
p^\mu= m  u^\mu  , \label{2.22}\end{eqnarray} where $m =\sqrt{p^\mu
p_{\mu}}$ is the inertial mass of the spinor. (\ref{2.22}) times
(\ref{220}),  we have
\begin{eqnarray}\frac 1 2 \frac{d}{d t}(p^{\mu}  p_{\mu})& =&msW_\ga v^\mu\pa_\mu G -m
W_F\frac d {dt} {\rm ln}\sqrt{1-v ^2},  \\
\frac{d}{d t}m &=& \frac d {dt}[sW_\ga G- W_F{\ln}(\sqrt{1-v
^2})],\label{2.23*}
\end{eqnarray} in which we used $\frac d {dt} = v^\mu \pa_\mu$.
Integrating (\ref{2.23*}) we get the moving inertial mass $m$ of the
spinor $\phi$
\begin{eqnarray}
m =m_0+sW_\ga G(t,\vec X)-W_F{\ln} {\sqrt{1-v^2 }},
\label{2.23}\end{eqnarray} where $m_0$ is constant static mass.
Substituting (\ref{2.23}) into (\ref{2.22}) we get
\begin{eqnarray}
p^{\mu} =\l(m_0+sW_\ga G-W_F{\ln} {\sqrt{1-v^2 }}\r)u^{\mu}.
 \label{2.24}\end{eqnarray}
Substituting (\ref{2.24}), (\ref{2.15}), (\ref{2.16}) and
(\ref{2.17}) into (\ref{2.8}), we finally get the complete
mass-energy relation for spinor $\phi$ as follows, which includes
the contributions of all self-potentials of the spinor.
\begin{eqnarray}
E~&=& E_0 -\frac{M_1 v^2}{\sqrt{1-v^2}}
+\frac{W_F}{\sqrt{1-v^2}}\ln\frac{1}{\sqrt{1-v^2}},\label{3.31}\\
E_0 &=& \frac {M_0}{\sqrt{1-v^2}},\label{3.32}\\
M_0 &=& m_0+W_F+sW_\ga G+W_G+W_A,\\
M_1 &=& W_F+\frac 1 3 (W_G-W_A)+W_b,
\end{eqnarray}
where $M_0$ is the total static mass of $\phi$. By (\ref{3.31}) and
(\ref{3.32}), we have
\begin{eqnarray}
\frac E{E_0}-1=\frac {W_F-2M_1}{2M_0}v^2+\frac {W_F}{4M_0}
v^4+O(v^6).\label{3.35}
\end{eqnarray}
(\ref{3.35}) can be tested by experiments, and 3 parameters
$(M_0,M_1,W_F)$ can be determined.

\section{Discussion and Conclusion}
\setcounter{equation}{0}

We derived the local Lorentz transformations for classical
parameters and established the relationship between field theory and
corresponding classical mechanics in a logical procedure. The
energy-speed relations for each potential term are derived, and the
classical mass of a particle is clearly defined. The subtle logic
relationship between concepts of quantum mechanics and that of
classical mechanics is overlooked previously and results in
confusions and puzzles. The above results imply the following
conclusions.

\begin{enumerate}
\item Classical mechanics once was a guidance to establish quantum mechanics. By analogy we define the concepts for quantum theory. However, from (\ref{2.1}) to (\ref{220}) and (\ref{2.24}), we learn that, the Dirac equation (\ref{2.3}) and field equations (\ref{2.4})-(\ref{219}) are more profound and general theory than the classical one (\ref{220}), because we can logically derive (\ref{220}) from (\ref{2.1}) under the assumption of mass-point model (\ref{*2.10}). Obviously, we cannot do the inverse operation. Analogy is not logic. When the more profound quantum theory goes to maturity, we should develop it relatively independent of classical mechanics. Otherwise, the overlapping concepts of two different systems certainly result in contradictions and confusions.

For spinor field $\phi$, we cannot specify its coordinate or
momentum without a clear definition, because $x^\mu$ is just a
presetting label system for space-time\cite{Ehren}, rather than the
coordinates of a particle. If we have not exact concepts for
coordinate and momentum of the spinor, what is the meaning of the
Heisenberg uncertainty relation? which is based on such concepts.
Obviously this situation is caused by the residual classical picture
of a particle in mind when we analyze the spinor field equation. As
a matter of factor, except for the fields $\phi$ and $A^\mu$ and so
on we have nothing else. $\phi$ and $A^\mu$ are basic properties for
a matter system, and other concepts should be logically defined from
them. In some cases a spinor acting like a particle is because
interactions and environment make it so, and in this time we should
clearly define classical concepts and then derive its simplified
Newtonian mechanics as well as the conditions of validity as done
above. Only in such procedure, confusions will be removed
automatically and a lot of debates are superfluous. For
(\ref{*2.8}), (\ref{*2.9}) and (\ref{2.24}), how does one use the
principle of uncertainty?

\item A more puzzling equation than (\ref{219}) is the Schr\"odinger equation for $N$-electron,
\begin{eqnarray}
\hbar i \pa_t \Psi=\hat {\textbf{H}} \Psi,\quad \hat
{\textbf{H}}=\sum^N_{k=1}\left(\frac{\hat p^2_k}{2m}+eV(t,\vec X_k)
+\frac{e^2}{8\pi }\sum_{l\ne k}\frac{1}
 {|\vec{X}_l-\vec{X}_k|}\right).
 \label{sdg}\end{eqnarray}
In $\hat{\textbf{H}}$ the coordinate $\vec X_k$ is quite similar to
the classical one for mass-point, and we have $3N$ coordinates for
$N$ electrons. The detailed derivation and  explanation for this
equation and the origin of the $3N$ coordinates is complex, which
were given in \cite{gu1,schr}. We do not repeat it here. What we
stress here is that, (\ref{sdg}) is also the non-relativistic
approximation of some juxtaposing Dirac equations. To explain the
foundation of quantum mechanics, pure philosophical speculation is
little helpful to solve the puzzles. Only if the logical structure
between the concepts and equations is clarified, the puzzles
automatically vanish.

\item Manifestly, field theories like (\ref{2.1}) are compatible with general relativity. If we take the quantum theory to be essentially a field theory, and QED or QCD to be just an inconvenient and ambiguous computing procedure to solve field equation, then consistency problem vanishes immediately.  Dirac equations have abundant solution spectra which could be solved only by reliable and systematic procedure constructed according to normal mathematics\cite{grn, prc, mass}.

Furthermore, along this line of opinions, we naturally get a kind of
unified field theory as follows\cite{quat}:

{\bf A1.} {\em The space-time is described by
\begin{eqnarray}
d\mathbf{x}=\wt\ga_\mu dx^\mu=\ga_\al \dl X^\al,
\label{1.1a}\end{eqnarray} in which  $\ga_\al$ and $\wt\ga_\mu$
satisfy the following $C\ell({1,3})$ Clifford algebra,
\begin{eqnarray}
\ga_\al\ga_\be+\ga_\be\ga_\al=2\eta_{\al\be},\qquad
\wt\ga_\mu\wt\ga_\nu+\wt\ga_\nu\wt\ga_\mu=2g_{\mu\nu}. \label{1.2a}
\end{eqnarray}}
{\bf A2.} {\em The dynamics for a definite physical system is given
by
\begin{eqnarray}\pa \Psi ={\cal F}(\Psi), \qquad \pa \equiv \wt \ga^\mu \pa_\mu, \label{1.5a}
\end{eqnarray}
in which $\Psi=(\psi_1,\psi_2,\cdots,\psi_n)^T$, and ${\cal
F}(\Psi)$ consists of some tensorial products of $\Psi$, so that the
total equation is covariant. }

{\bf A3.} {\em The solutions to (\ref{1.5a}) are singularity-free,
that is, $\forall \psi_k\in L^\infty$.}

The Dirac equation, Maxwell equation and  Einstein equation all
satisfy (\ref{1.5a}). By classifying $\Psi$ according to spin
indices $s=\frac1 2, 1,2$ we can derive these equations respectively
as well as their coupling system. Under the constraint of {\bf A3},
we have only limited choices for the interactive potentials. If we
solve the dynamics (\ref{1.5a}) from the simplest system to more
complicated ones, we will certainly reach the right results for all
physical problems. In the standard model, the constraints of
assumptions such as $SU(2)$ and $SU(3)$ are too strong and too
narrow to contain enough physical objects, and the theory is trapped
in complicated formalism. The more the constraints, the larger the
risk and the less the truth.

\item In some textbooks, the relations (\ref{*2.14})-(\ref{*2.16}) are directly derived via Lorentz transformation of the integrands and volume element relation $ d^3x=\sqrt{1-v^2}d^3 \bar x$. This calculation is only an approximation in general cases, because parameters in (\ref{*2.16}) depend on $\bar t$ in this case. It is accurate only for spinor at particle state. However, it is usually accurate enough for particles such as electron and proton due to the static mass energy much larger then energy exchanged with the environment.

Usually, the proper parameters have very simple form.  For  true
vector, it usually takes the form $\bar I^\mu=(\bar I^0,0,0,0)$,
then we get
\begin{eqnarray}
I^{\mu}=\sqrt{1-v^2}\bar I^0 u^{\mu}.\label{*2.21} \end{eqnarray} In
some cases, the symmetrical true tensor $ I^{\mu\nu}$ is given by
\begin{eqnarray}
I^{\mu\nu}=\sqrt{1-v^2}\left(\bar K u^\mu u^\nu+\bar
J\eta^{\mu\nu}\right),\label{*2.22} \end{eqnarray} where $\bar K,
\bar J$ are constants.

\item By (\ref{2.24}), (\ref{2.15}), (\ref{2.16}) and (\ref{2.17}), we learn different potential has different energy-speed relation, so we can identify the interactive potentials by testing the fine structure of the energy-speed relation of a particle\cite{tst}. Especially, the existence of nonlinear potential $F$ can be determined, which is much important to disclose the nature of fundamental particles and dark matter. The linear Dirac equation is accurate for Hydrogen atom spectrum. How an electron balances its self-electromagnetic interaction can also be discovered by testing the mass-energy relation of an electron\cite{tst}.

However, for normal particles such as electron, the numerical
results\cite{3}-\cite{gu3} showed that the mass contributed by
potentials $(W_F, W_A, W_G)$ and so on are much less than the static
mass $m_0\approx \mu$, so Einstein's mass-energy relation holds to
high precision if $v\ll c$.

\item By (\ref{2.23}) or (\ref{2.24}) we find the external scalar $G$ manifestly influence the inertial mass and
momentum of a particle. This effect will seriously violate classical mechanics, so such a scalar field should be absent in Nature.

\end{enumerate}

\section*{Acknowledgments}
The author is grateful to his supervisor Prof. Ta-Tsien Li and Prof.
Han-Ji Shang for their encouragement. It is my pleasure to thank the
reviewer A for helping to clarify some subtle concepts and to
correct a number of language errors.


\begin{thebibliography}{99}
\bibitem{E1} Einstein, A. Ann. der Phys. 18, 639-641 (1905).
\bibitem{gu1} Y. Q. Gu, {\em New Approach to N-body Relativistic
Quantum Mechanics}, Int. J. Mod. Phys. A22:2007-2020(2007),
arXiv:hep-th/0610153
\bibitem{sml} Lee Smolin, {\em The truble with physics}, Spin Network, Ltd. 2006.
\bibitem{3} R. Finkelsten, et al, Phys. Rev. {\bf 83(2)}, 326-332(1951)
\bibitem{4} M. Soler, Phys. Rev. {\bf D1(10)}, 2766-2767(1970)
\bibitem{5} M. Wakano, Prog. Theor. Phys. {\bf 35}, 1117(1966)
\bibitem{6} M. Soler, Phys. Rev. {\bf D8}, 3424(1973)
\bibitem{7} A. F. Randa, M. Soler,  Phys. Rev. {\bf D8}, 3430(1973)
\bibitem{gu3} Y. Q. Gu, {\em Some Properties of the Spinor
Soliton}, Adv in Appl. Clif. Alg. {\bf V8(1)}, 17-29(1998)
\bibitem{*gu7} Y. Q. Gu, {\em A Cosmological Model with Dark Spinor Source},
Int. J. Mod. Phys. A22:4667-4678(2007), gr-qc/0610147
\bibitem{sbt} Y. Q. Gu, {\em Some Subtle Concepts in Fundamental Physics}, to appear in PHYSICS
ESSAYS, arXiv:0901.0309
\bibitem{quat}  Y. Q. Gu, {\em The Quaternion Structure of Space-time and Arrow of
Time},  J. Mod. Phys. {\bf V3}, 570-580(2012).
\bibitem{grn} W. Greiner, B. M\"uller, {\em Quantum Mechanics -- Symmetries}, Springer, 1994.
\bibitem{Ehren} Y. Q. Gu, {\em Some Paradoxes in Special Relativity
 and resolutions}, Adv in Appl. Clif. Alg. V21(1), 103-119(2011),  arXiv:0902.2032
\bibitem{schr}  Y. Q. Gu, {\em The Electromagnetic Potential Among Nonrelativistic
Electrons}, Adv. Appl.  Cliff. Alg. {\bf 9(1)}, 61-79(1999)
\bibitem{prc}  Y. Q. Gu, {\em A Procedure to Solve the Eigen Solution to Dirac Equation}, Quant. Phys. Lett. 6, No. 3,161-163 (2017);, arXiv:0708.2962
\bibitem{mass}  Y. Q. Gu, {\em Mass Spectrum of Dirac Equation with Local Parabolic Potential},\\ arXiv:hep-th/0612214
\bibitem{tst} Y. Q. Gu,  {\em  Test of Einstein's Mass-Energy Relation},
arXiv:hep-th/0610189v3

\end{thebibliography}
\end{document}